\documentclass[aps,pre,twocolumn]{revtex4-1}
\usepackage{amsmath}
\usepackage{graphicx}
\usepackage{bm}
\usepackage{amssymb}
\usepackage{array}
\usepackage{mypack}

\newcommand{\calK}{{\cal K}}
\newcommand{\calL}{{\cal L}}

\graphicspath{{figures/}}

\begin{document}
\title{Revisiting the entropic force between fluctuating biological membranes}
\author{Y. Hanlumyuang$^1$, L.P. Liu$^{2}$, P. Sharma$^{1,3}$$^\clubsuit$ \\
{\em $^1$ Department of Mechanical Engineering, University of Houston, TX, USA}\\
{\em $^2$ Department of Mechanical \& Aerospace Engineering and Department of Mathematics}\\
{\em Rutgers University, Piscataway, N.J., U.S.A.}\\
{\em $^3$ Department of Physics}\\
{\em University of Houston, TX, U.S.A.}
}
\begin{abstract}
The complex interplay between the various attractive and repulsive forces that mediate between biological membranes governs an astounding array of biological functions: cell adhesion, membrane fusion, self-assembly, binding-unbinding transition among others. In this work, the entropic repulsive force between membranes---which originates due to thermally excited fluctuations---is critically reexamined both analytically and through systematic Monte Carlo simulations. A recent work by Freund \cite {Freund13} has questioned the validity of a well-accepted result derived by Helfrich \cite{Helfrich78}.  We find that, in agreement with Freund, for small inter-membrane separations ($d$),  the entropic pressure scales as $p\sim 1/d $, in contrast to Helfrich's result: $p\sim 1/d^3$.  For intermediate separations, our calculations agree with that of Helfrich and finally, for large inter-membrane separations, we observe an exponentially decaying behavior.
\end{abstract}
\maketitle

\section{Introduction}
Biomembranes exert various types of repulsive and attractive forces on each other. The van der Waals forces are weakly attractive---these vary as $1/d^3$ for close separations and transition to $1/d^5$ at larger distances \cite {Helfrich84, Ninham70}. Here $d$ is the mean distance between the interacting membranes. The notable aspect of the attractive force is that it is long-ranged.  A somewhat ambiguous term \emph{hydration forces} \cite {Rand89}, is used to denote the repulsive force that mediates at very small inter-membrane distances. The underlying mechanisms of hydration forces are still under active research \cite {Lipowskychapter}---it suffices to simply indicate here that they are quite short ranged and drop off exponentially with distance.

Helfrich, in a pioneering work \cite{Helfrich78}, showed that two fluctuating fluid membranes exert a repulsive force on each other. Biomembranes are generally quite flexible and a single membrane fluctuates both freely and appreciably at physiological temperatures. As two membranes approach each other, they \emph{hinder} or diminish each others out-of-plane fluctuations. This hindrance decreases the entropy and the ensuing overall increase of the free-energy of the membrane system, which depends on the inter-membrane distance, leads to a repulsive force that tends to push the membranes apart. Helfrich \cite {Helfrich78}, using a variety of physical arguments and approximations, postulated that the entropic force varies as $1/d^3$.  In contrast to the other known repulsive forces, this behavior is long-ranged and competes with the van der Waals attraction at all distances \cite {Helfrich84, Ninham70, Israel92, Milner92, Lipowsky86, Lipowsky89}. Since Helfrich's proposal \cite {Helfrich78}, biophysicists have used the existence of this repulsive force to explain and understand a variety of phenomena related to membrane interactions.  Helfrich's work has been reexamined and extended in Ref. \cite{Janke86, Gouliaev98, Kleinert99} (among others) and  most recently by Freund \cite{Freund13}. See also \cite {Sharma13} for a an overview of Freund's work.

Freund clearly highlights some of assumptions made in Helfrich's work and provides a fresh perspective on this problem \cite{Freund13}. Freund controversially finds that within a range of $d$ values the force law between two fluctuating membranes is proportional to $1/d$  rather than the well-accepted result of Helfrich: $1/d^3$. To settle this issue, we have reexamined this problem both analytically and through recourse to carefully conducted Monte Carlo simulations. As was initially pointed out by Helfrich \cite{Helfrich78}, due to reflective symmetry, the evaluation of the force between two membranes in a periodic stack may be replaced by a single membrane confined between two rigid walls (Figure 1). Throughout this article, we will emphasize the differences between our work and those of Ref. \cite{Freund13, Janke86,  Helfrich78}.

\begin{figure}
\centering
\includegraphics[scale=0.45,clip]{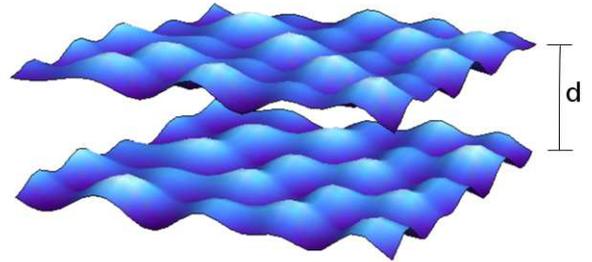}
\caption{\label{fig:1}
A pair of fluctuating membrane may be replaced by a single membrane confined between two walls separated from each other by a distance $2d$. }
\end{figure}

\section{General Formalism and Asymptotic Limit}

Consider a membrane as depicted in Fig. \ref{fig:1}. Assume that the membrane occupies $S=[0, L]^2$ on the $x y $-plane and the thermodynamic state of the membrane is described by $u\equiv u_z(x ,y )$, where $u_z$ is membrane mid-plane deviation along the $z$-axis. In the Helfrich model, the Hamiltonian is
\begin{equation}
H[u] = \int_S \dd ^2\xv \frac{\kappa}{2}(\partial^2u)^2.
\label{eqn:1}
\end{equation}

To address the thermal fluctuation we make the following assumptions:
(i) the membrane consists of  $N$ molecules located at $\xv\in \calL=\{a \left(l_1, l_2\right): l_1, l_2=1,\cdots, m\}$, where $a=L/m$ is the molecule's size. In other words, $N=m^2$ is the total degrees of freedom of the system, and (ii) microscopically  the out-of-plane deviation $ u_{\xv}$ of each molecules is quantized and can only take values from $\{ n\delta d: n=1,\cdots, d/\delta d \}$, where $\delta d$ is a small spacing  along the deviation direction. Then from the definition \cite{Kittel80} the partition function of the system can be written as a functional integral:
\begin{equation}
Z=\int_{-d}^{d}\prod_{\xv\in \calL}C\dd u_{\xv} e^{-\beta\int_S  \dd ^2\xv \frac{\kappa}{2}(\partial^2u)^2},
\label{eqn:2}
\end{equation}
where $\beta=1/k_B T$ and $u=u(\xv)$ is any differentiable function interpolating the discrete molecules' deviation $u_{\xv}$, $\xv\in \calL$.

The point-wise hindrance condition that $|u_{\xv}|\le d$ is enforced throughout this work. This constraint is in fact the key obstacle in the closed-form evaluation of the partition function.
Freund \cite{Freund13} modified the partition function by introducing
adjustable integration limits, and then minimized the resultant free energy with respect to these limits to determine the change in free energy with respect to inter membrane separation (and hence the entropic force law). In the Conclusions section, we discuss Freund's approach further. Here, we adopt the form of the partition function in Eq.(\ref{eqn:2}) which, notwithstanding the analytical intractability of its integration limits (i.e. pointwise constraint on $u$), is \emph {exact} within  the present formulation. We remark here that Helfrich \cite{Helfrich78} avoids the point-wise $hindrance$ condition, $\left| u(x,y)\right|\leq d$, and instead imposes a weaker constraint where
the mean square membrane displacement is required to be bounded by $d^2$, i.e.  $\langle u^2 \rangle\leq d^2$.

\newcommand{\vtld}{\tilde{v}}
\newcommand{\Zbar}{\bar{Z}}
\newcommand{\sqrttau}{\sqrt{\tau}}
To gain new insights into the partition function,  two dimensionless quantities, $y$ and $v$, are introduced as
\begin{equation}
 \yv = \frac{\xv}{L}\quad\text{and}\quad v(\yv)= \frac{u(\xv)}{d}.
 \label{eqn:3}
 \end{equation}
By changes of variables we may rewrite the partition function \eqref{eqn:2} as ($S_0=[0,1]^2$, $A=L^2$)
\begin{equation}
Z =  (Cd)^N\int_{-1}^{1}\prod_{\yv \in \tilde{\calL} } \dd v_{\yv}
 e^{-\frac{\beta \kappa d^2 } {2 A } \int_{S_0}\dd ^2\yv (\partial^2_{\yv} v)^2} ,
 \label{eqn:4}
\end{equation}
where  $\yv\in \tilde{\calL}=\{\left(l_1/m, l_2/m\right): l_1, l_2=1,\cdots, m\}$.
Let $\tau \equiv A/\beta\kappa d^2$ be a dimensionless variable. By a change of variable back and forth, 
$v /\sqrt{\tau} \leftrightarrow\  \vtld  $, we obtain
\begin{equation}
Z = (Cd)^N\left(\sqrt{\frac{A}{\beta\kappa d^2}}\right)^N \bar{Z}(\tau),
\label{eqn:5}
\end{equation}
where
\begin{equation}
\bar{Z}(\tau) = \tau^{-N/2}\int_{-1}^{1}\prod_{\yv \in\tilde{\calL}} \dd v_\yv  e^{-\frac{1}{2\tau} \int_{S_0}\dd^2\yv(\partial^2_\yv v)^2}.
\label{eqn:6}
\end{equation}
As a consequence, the free energy density has the form
\begin{equation}
\Ff = -\frac{k_{B}T}{2 a^2 }\ln\left( \frac{k_B T A C^2}{\kappa }\right) -
\frac{k_{B}T}{ A } \ln \bar{Z}\left(\frac{k_B T A}{\kappa d^2}\right).
\label{eqn:7}
\end{equation}
 It should be noted
here that our free energy density  has a subtle difference from the one in Ref. \cite{Janke86}. The first term in the free energy does not depend on $d$ and vanishes upon differentiation with respect to $d$.
It follows that the steric pressure is
\begin{equation}
p = -\frac{1}{2}\frac{\partial \Ff}{\partial d} = \frac{(k_BT)^2}{ \kappa d^3 }g(\tau).
\label{eqn:8}
\end{equation}
\noindent
where $g(\tau)  \equiv - \partial \ln \bar{Z}(\tau)/ \partial \tau$.
Janke and Kleinert \cite {Janke86} have performed Monte Carlo calculations and found that $g(\tau)$ is a constant in the thermodynamic limit \cite{Janke86}.
The value of this constant was also found  by Kleinert using a variational approach \cite{Kleinert99}. We remark that in the past works that have performed Monte Carlo calculations of this problem, the veracity of the Helfrich's pressure law has been implicitly embraced and the focus has not been on examining the dependence of the entropic pressure on inter membrane distance but rather calculation of $g(\tau)$ \emph{assuming} that Helfrich's inverse cube law is correct. This is the reason, we believe, that Freund's result and (now ours) has not been noted until now.

As evident from the definition in Eq.(\ref{eqn:8}), $g (\tau)$ is a 
function of the separation between the rigid walls confining the membrane. 
It is interesting to consider the energy variation, and consequently the pressure, 
at some small distances. As Freund has shown, this limit is analytically tractable. 
Below we reproduce the result using a slightly different procedure.  
Consider the scaled partition function in Eq.(\ref{eqn:6}),  by Fourier transformation we introduce
\begin{equation}
\begin{split}
\vh_{\kv} =\frac{1}{m}\sum_{\yv\in \tilde{\calL}}e^{-i\kv\cdot\yv}v_{\yv}, \qquad v_{\xv}  =\frac{1}{m}\sum_{\kv\in  \tilde{\calK }} e^{i\kv\cdot\xv}\vh_{\kv},
\end{split}
\label{eqn:10}
\end{equation}
\noindent
where  $ \tilde{\calK}=\{2\pi\left(n_1, n_2\right): n_1, n_2=1,\cdots, m\}$ is the reciprocal lattice of $\tilde{\calL}$. In matrix notations the above equations can be rewritten as
\begin{equation}
\vec{v}_\yv = \Uu^\dagger  \vec{\vh}_\kv, \qquad \vec{\vh}_\kv = \Uu \vec{v}_\yv,
\label{eqn:11}
\end{equation}
where $\vec{v}_\yv$ (resp. $\vec{\vh}_\kv$) denotes the column vector formed by $v_{\yv}$, $\yv\in \tilde{\calL}$ (resp.  $\vh_{\kv}$, $\kv\in \tilde{\calK} $), and $\Uu$ is a unitary matrix satisfying $\Uu^\dagger \Uu = \Uu\Uu^\dagger = 1$. Since $a<<L$, we may convert  an integral over $S$ as a summation over $\calL$: $\int_S\dd^2\xv\Rightarrow a^2\sum_{\xv\in \calL}$, and consequently $  A \int_{S_0}\dd^2\yv \Rightarrow a^2 \sum_{\yv \in \tilde{\calL} }$ . Applying the Parseval's theorem, 
we rewrite
\begin{equation}
\begin{split}
\int_{S_0}\dd^2\yv(\partial^2_y v )^2 & = \frac{1}{m^2}\sum_{\kv\in \tilde{\calK} } |\vh_\kv|^2 |\kv|^4\\
&=\frac{1}{m^2}\vec{v}_\yv \cdot\Uu^{\dagger}\Dd(\kv) \Uu\vec{v}_\yv,
\end{split}
\label{eqn:12}
\end{equation}
where  $\Dd(\kv) $ is the diagonal matrix with entries $|\kv|^4$, $\kv\in \calK$.  
Defining a dimensionless variable $ \pv \equiv \kv/m^4 $, the scaled partition function is 
\begin{equation}
\begin{split}
\bar{Z}(\tau) =  \tau^{-N/2} \int_{-1}^{1}\prod_{\yv\in \tilde{\calL}   } C\dd v_\yv
e^{-\frac{m^2}{2\tau}  \vec{v}_\yv \cdot\Uu^{\dagger}\Dd(\pv) \Uu\vec{v}_\yv}
\end{split}
\label{eqn:13}
\end{equation}
%
For $\tau\rightarrow \infty$  i.e. $d/a\rightarrow 0$,  one has
\begin{equation}
\begin{split}
\bar{Z}(\tau) =& \tau^{-N/2}\left[2^N - \frac{m^2}{2\tau} (\Uu^\dagger \Dd\Uu)\cdot\int_{-1}^{1}  \prod_{\yv\in\tilde{\calL}}\dd v_{\yv}
\vec{v}_\yv\otimes\vec{v}_{\yv} \right.\\
&\quad\quad \quad\quad\left.+\Oo(\frac{d^4}{a^4})
 \right],
 \end{split}
 \label{eqn:14}
\end{equation}
where the identity $\vec{v}_\yv\cdot\Uu^\dagger\Dd\Uu\vec{v}_{\yv} = (\Uu^\dagger \Dd\Uu)
\cdot(\vec{v}_\yv\otimes \vec{v}_{\yv})$ was used. 
Since $\int_{-1}^{1}\prod_{\yv\in \tilde{\calL} } \dd v_{\yv}\vec{v}_\yv\otimes \vec{v}_{\yv}
= \frac{2}{3} 2^{(N-1)}\Ii = \frac{2^N}{3}\Ii$, and the inner product  of $\Uu^\dagger\Dd\Uu $
with  the identity matrix yields its trace, the reduced partition function in the asymptotic limit is
\begin{equation}
\begin{split}
\bar{Z}(\tau) =& \tau^{-N/2}\left[2^N- \frac{2^N}{3} \frac{m^2}{2\tau}
\sum_{m^4p\in \tilde{\Kk}} p^4 + \Oo(\frac{d^4}{a^4}) \right].
\end{split}
\label{eqn:15}
\end{equation}
It is clear from the definition $g(\tau) = -\partial \ln \bar{Z}/\partial \tau $  that the steric pressure
has the leading order term of $p\approx k_BT/2da^2$. The next correction term can be obtained
using the identity $\sum_{m^4 p\in\tilde{\Kk} } p^4 = (2\pi)^4 \left[ 23 m^2/45 - 26 m/15 + \Oo(1) \right]$,  where $m^2 = N$.
The steric pressure in the limit $\tau\rightarrow \infty$ ($d/a\rightarrow 0$) is thus
\begin{equation}
p = -\frac{1}{2}\frac{\partial \Ff}{\partial d}
= \frac{k_BT}{2a^2d}\left[ 1 - (2\pi)^4 \frac{46}{45} \frac{\beta \kappa d^2}{6a^2} +
\Oo(\frac{d^4}{a^4}) \right].
\label{eqn:16}
\end{equation}
\noindent
This pressure law has a very different $d-$dependence 
than any known theories or simulations to date 
\cite{Helfrich78, Janke86, Kroll89, David90, Netz95, Gouliaev98, Kleinert99} 
except of course the work by Freund \cite{Freund13}.

Another interesting point obtained from Eq.(\ref{eqn:16}) is that the first term in the right-hand side has the form of pressure of the ideal gas.  Physically, this implies that the membrane fluctuates similar to an ideal gas in the limit $d/a\rightarrow 0$, with a correction term which is in the order of $\Oo(d)$ (the second term in Eq.(\ref{eqn:16})). This ideal-gas contribution does not depend on the bending modulus $\kappa$, hence every type of membranes should exhibit the $p \sim 1/d$ pressure law at the limit $d/a\rightarrow 0$. 

In order to elucidate the full $d-$dependence including the limit $d/a\rightarrow 0$, a natural step is to take recourse in numerical Monte Carlo simulations which are discussed in the next section.

\section{Monte Carlo Simulations}
In the Monte Carlo simulations, the spatial coordinates are replaced by a square grid  $\{\xv\}$ with lattice constant $a$. 
The membrane displacement along the $z-$direction is specified by $u_\xv \equiv u(\xv)$, where 
it is also discretized to a grid of space $\delta d$. This scheme has been shown to be sufficient 
in prior Monte Carlo calculations \cite{Janke86, Lipowsky89}. The Hamiltonian over these lattice points is
\begin{equation}
H  = a^2\sum_{x\in\calL}\frac{\kappa}{2}(\partial^2u)_\xv^2,
\label{eqn:17}
\end{equation}
\noindent
where the discretized Laplacian
\begin{equation}
(\partial^2u)_\xv = \frac{1}{a^2}\left[\sum_{\hat{\rho}\in \text{nbr}} u_{\xv+\hat{\rho}}- 4u_{\xv}\right],
\label{eqn:18}
\end{equation}
and $\hat{\rho}$ denotes the displacement vectors to the four nearest neighbors
to the site $\xv$.

\begin{figure}
\centering
$\begin{array}{c}
\includegraphics[scale=0.45,clip]{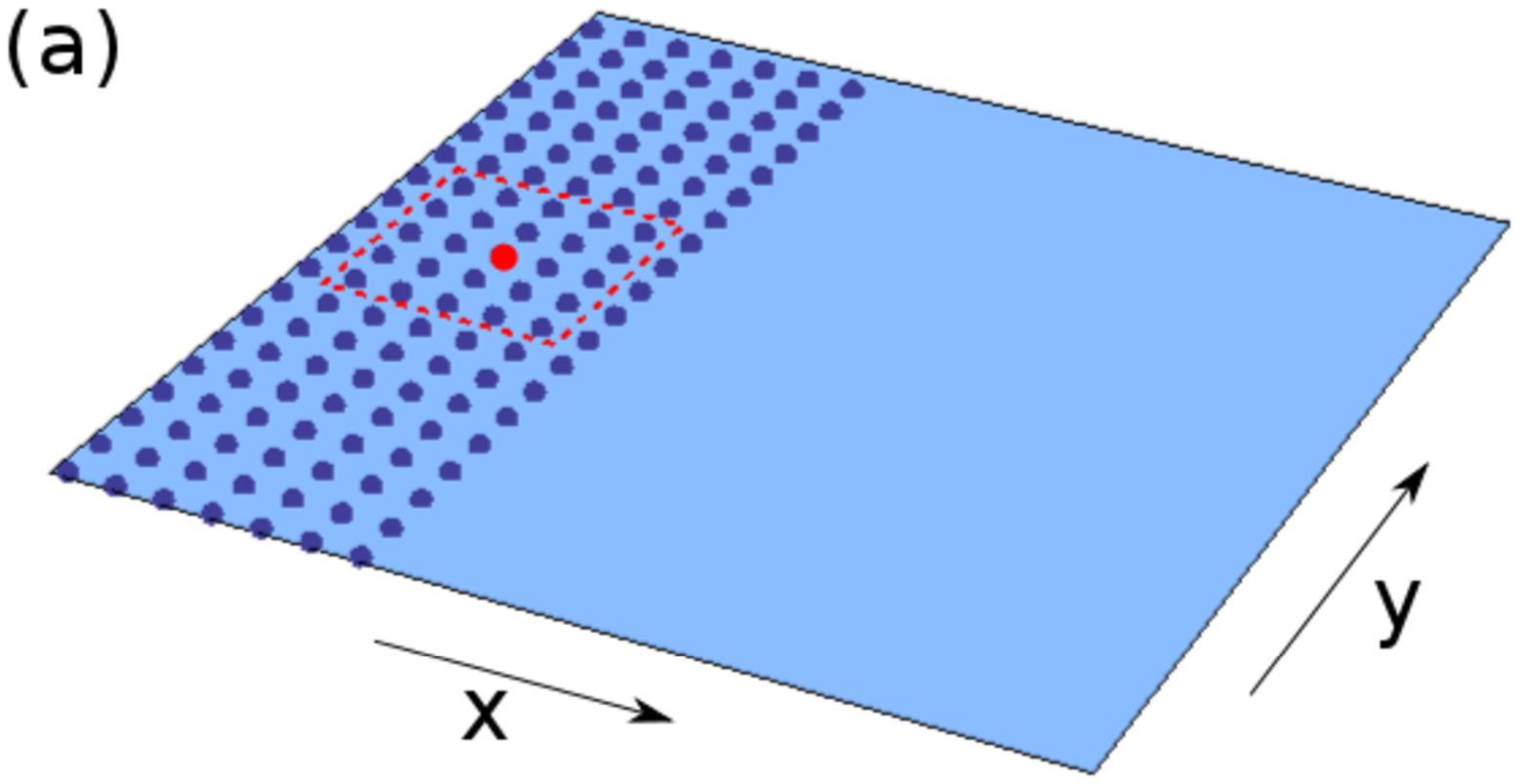} \\
\includegraphics[scale=0.45, clip]{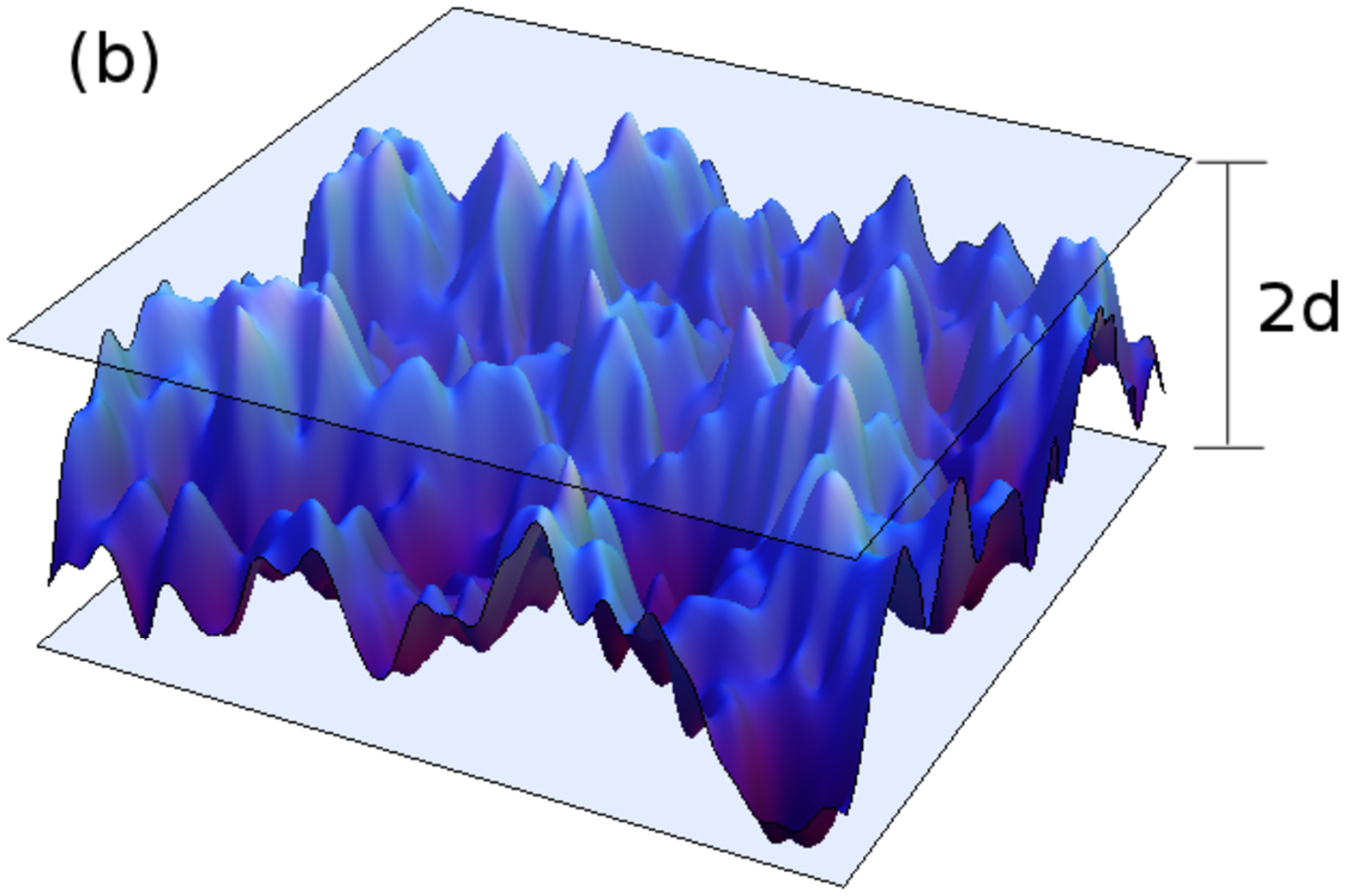}\\
\end{array}
$
\caption{\label{fig:2}  (a)  Discretization of a strip geometry of a membrane.  To update the central point (red dot), the knowledge of the neighboring points  inside the
dashed square is required.  The $x-$direction of the strip is treated within the  realm of message passing, while a periodic boundary 
condition is employed along the $y-$direction.  (b) A realization of a membrane of size 40 by 40 from an MC run. The energetic parameters
are $\kappa = 1.0$, $k_BT = 2.0$, while for the length scales $a = 1.0$ and $d=5.0$. }
\end{figure}
Our simulation code was fully parallelized using spatial decompositions. The lattice was divided into strips so that each strip can 
be updated independently using the usual Metropolis algorithm  \cite{Nakano, Heermann91}. Since each strip is updated 
in parallel, some care has been taken in order to account for the lattice sites at the boundary. Figure \ref{fig:2}(a) illustrates
a typical geometry in the simulations.  Updating the central point requires the knowledge of the values at all points inside the dashed
square.  It is forbidden to update any
other site in this neighborhood until the update of the central site has been completed. This can be accomplished 
by an appropriate choice of strip lengths along the $x-$axis and performing a row-by-row update.  In our simulations, 
the minimal lattice length along the $x-$axis is 5, so that there is no overlap between dashed
square in Figure \ref{fig:2}(a). A message passing routine was employed in updating the top and bottom two rows of the strips, 
while along  the $y-$axis a simple periodic boundary condition was used.  In this way, our choice of lattice sizes are 50 by 50,  100 by 100, 600 by 600, and 1000 by 1000.
An MC realization is shown in Figure \ref{fig:2}(b), where the parameters used for its generation are listed in the caption.

In addition to the ensemble average of the energy  $\bar{\Ee}$, the second physical quantity needed for this problem is the pressure. The pressure can be derived by differentiating the change in free energy with respect to the inter membrane separation. Alternatively and more conveniently, the pressure can be related
to the ensemble average of the Hamiltonian density $\Hh = H/ Na^2$. 
Rewriting Eq.(\ref{eqn:4}) as
\begin{equation}
Z = (Cd)^N\int_{-1}^{1}\prod_{\yv\in\tilde{\calL}} \dd v_{\yv} e^{-\frac{1}{ \tau} \tilde{H}(\{ v_\yv \}) }
\label{eqn:19}
\end{equation}
where $\tilde{H} = (1/2)\int_{S_0}\dd^2\yv (\partial^2_\yv v_\yv)^2 \equiv  \tau \beta H $ is a scaled  Hamiltonian. The membrane 
configuration $\{u_\xv\}$ is parameterized by the set of variable $\{v_\yv = u_{\xv}/ d\}$.
Differentiation of the free energy  can be carried out straightforwardly by using these
rescaled parameters $\{ v_{\yv} \}  $; the concept which also appears in the derivation for
 the Hellmann-Feynman forces in quantum mechanics \cite{Feynman39}. It follows that
\begin{equation}
\begin{split}
\frac{\partial F}{\partial d}  =\frac{-k_BT}{Z} &
\left[
(Cd)^N \int_{-1}^{1} \prod_{\yv \in \tilde{\calL}} \dd v_{\yv} \left[-\frac{2 \tilde{H}(\{v_{\yv}\})}{d\tau} \right] e^{-\frac{1}{\tau} \tilde{H}(\{v_{\yv}\})}\right.\\
&\left.
+\frac{N}{d} (Cd)^N  \int_{-1}^{1} \prod_{\yv\in\tilde{\calL}} \dd v_{\yv} e^{-\frac{1}{\tau} \tilde{H}(\{v_\yv\} )} \right]
\end{split}
\label{eqn:20}
\end{equation}
Using the concept of ensemble average and Eq.(\ref{eqn:19}), the derivative
of the free energy is
\begin{equation}
\frac{\partial F}{\partial d} = \frac{2}{d}\ave{H} - \frac{N k_B T}{d}
\label{eqn:21}
\end{equation}
In term of the energy density $\Ff = F/ A$  and  $\Hh = H/A$ for a continuos membrane, or $\Ff = F/Na^2$ and $\Hh = H/Na^2$ for a discretized membrane, where $N$ is the number of molecules and $a^2$ is a square encompassing each of them
\begin{equation}
\frac{\partial \Ff}{\partial d} = \frac{2}{d}\ave{\Hh} - \frac{Nk_BT}{L^2 d }
= \frac{2}{d}\ave{\Hh} - \frac{k_BT}{a^2d}.
\label{eqn:22}
\end{equation}
Consequently for a membrane situated between two rigid plates of separation $2d$, the steric
pressure
\begin{equation}
p = -\frac{\partial \Ff}{\partial (2d) }  =  \frac{k_BT}{2a^2d} - \frac{1}{d}\ave{\Hh} = \frac{k_BT}{2a^2d} - \frac{1}{d}\bar{\Ee},
\label{eqn:23}
\end{equation}
where $\bar{\Ee} = \ave{\Hh}$.
The above pressure is exact, and makes the computation of its value straightforward.   Furthermore, the errors on
the pressure $p$ stems from only the errors on $\ave{\Hh}$, which for sufficiently long Monte Carlo 
time steps become small compared to the value of $\ave{\Hh}$. 

The validity of Eq.(\ref{eqn:23}) for both the asymptotic and general form of the pressure can
be easily checked. In the asymptotic limit $\tau\rightarrow\infty$ or $d/a\rightarrow 0$,
Eq.(\ref{eqn:15}) gives the free energy density $\Ff  = -k_B T \ln Z/Na^2 $ of the form
\begin{equation}
\Ff(\tau) =  -\frac{k_BT}{a^2}\left[  \ln(2dC) - (2\pi)^4\frac{23}{45}\frac{ k_B T}{a^2}\frac{m^2}{2\tau} +\Oo(\frac{d^4}{a^4})
 \right]. 
 \label{eqn:24}
\end{equation}
It then follows that
\begin{equation}
\bar{\Ee}  = \frac{\partial}{\partial \beta} \beta \Ff \approx (2\pi)^4 \frac{23}{45}\frac{\kappa d^2}{6 a^4}.
\label{eqn:25}
\end{equation}
\noindent
and 
\begin{equation}
p \approx \frac{k_BT}{2a^2d} -  (2\pi)^4 \frac{23}{45}\frac{\kappa d}{6 a^4},
\label{eqn:26}
\end{equation}
in agreement with Eq.(\ref{eqn:16}). In the general case, it can be deduced from Eq.(\ref{eqn:7}) that
\begin{equation}
\bar{\Ee}
= \frac{k_BT}{2a^2} - \frac{(k_BT)^2}{\kappa d^2}g(\tau).
\label{eqn:27}
\end{equation}
As a result, the pressure obtained from Eq.(\ref{eqn:23}) is
\begin{equation}
p = \frac{(k_BT)^2}{\kappa d^3}g(\tau),
\label{eqn:28}
\end{equation}
\noindent
which is again in agreement with Eq.(\ref{eqn:8}). 

\begin{figure}
\includegraphics[scale=0.4, clip]{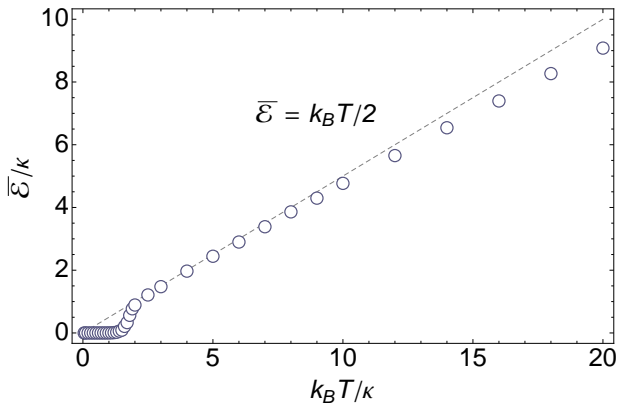}
\caption{\label{fig:3} A typical heating of a confined membrane. Circles represent MC results, while
the dashed line corresponds to the heating of one-dimensional ideal gas. The rigid wall separation
is $d=5.0$, while the membrane discretized spacing is $a = 1.0$.  The bending modulus is $\kappa = 1.0$.
The membrane starts to experience the presence of the walls at about $k_B T > 10\kappa $.
The errors of the energy is about $\pm 0.01$ . For example, the energy
on the far right point is $\bar{\Ee} = 9.080\pm 0.009. $ }
\end{figure}

\begin{figure}
$\begin{array}{c}
\includegraphics[scale=0.5, clip]{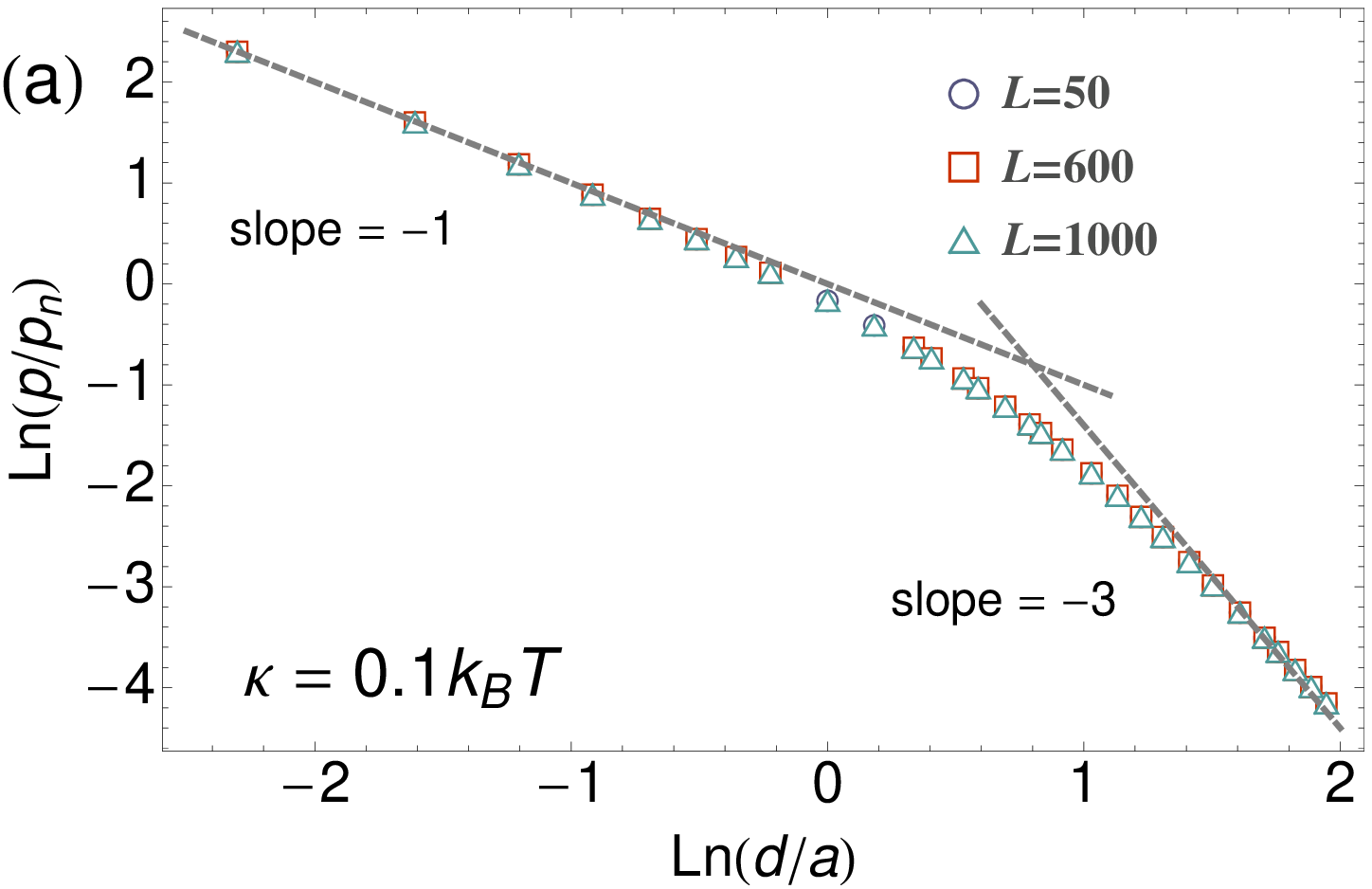}\\
\includegraphics[scale=0.5, clip]{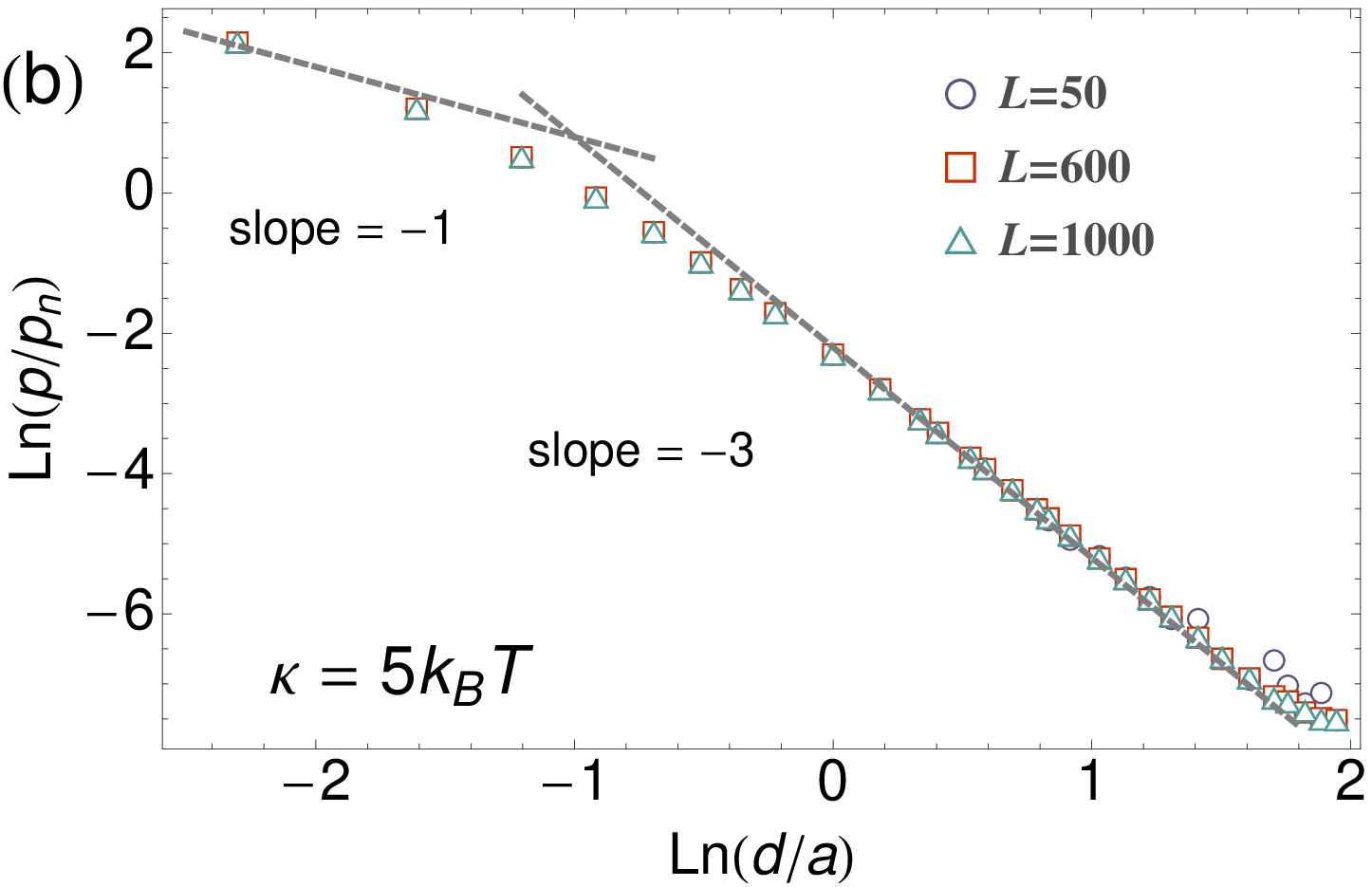}\\
\includegraphics[scale=0.5, clip]{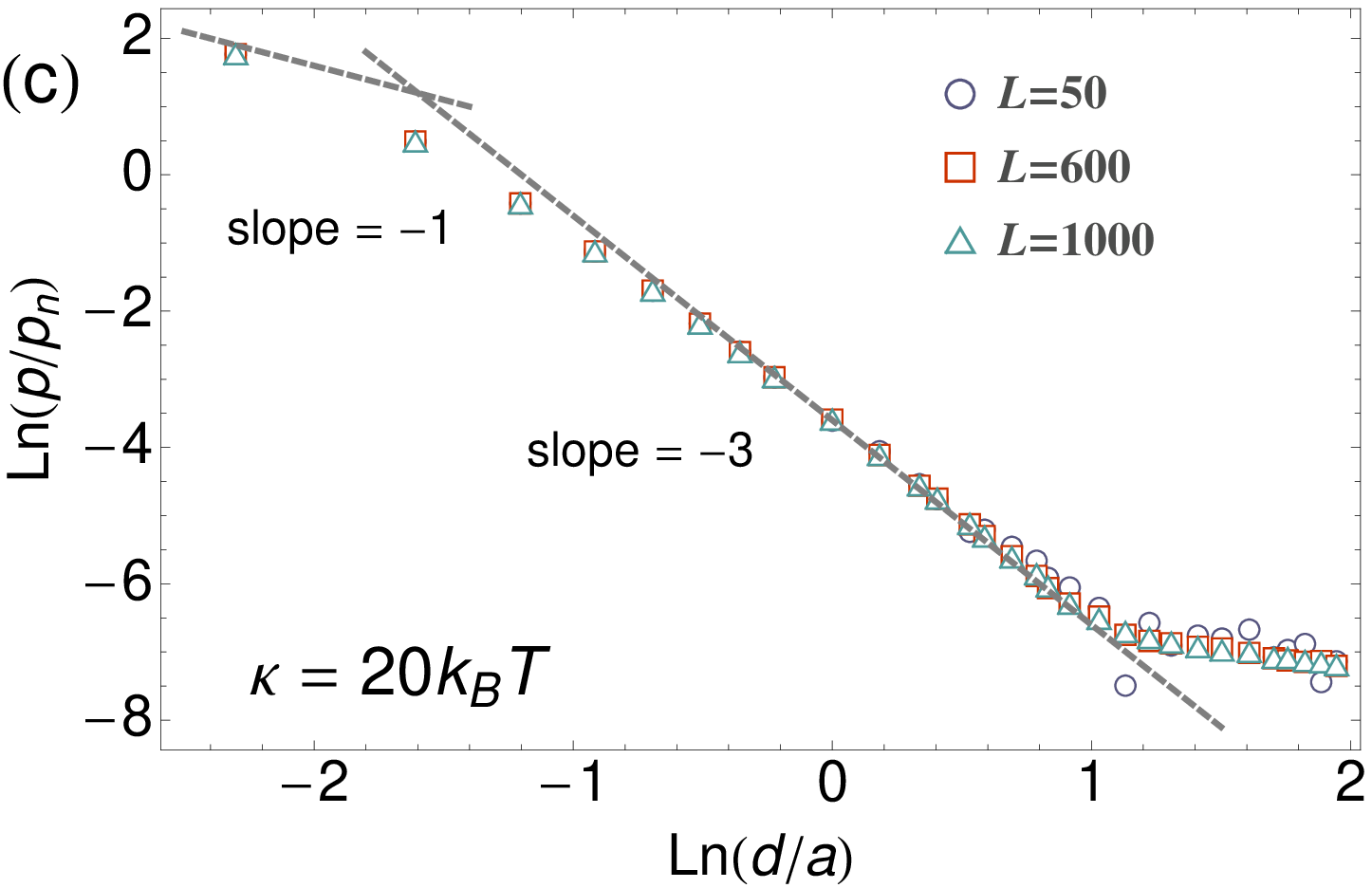}\\
\end{array}
$
\caption{\label{fig:4}Simulation results  (symbols) of the pressure $p$ and the distance $d$ between two rigid walls in  log-log formats. The lines are  drawn to guide the eye . 
In subfigures (a), (b), and (c) the bending modulus of $\kappa = 0.1 k_BT$, $5 k_BT$ and $20 k_BT$ are used respectively, whereas in all subfigures $\delta d = 0.1$, $a=1.0$, and $k_BT =10.0$.  For each bending modulus $\kappa$, the  sizes of the membranes $L\times L$  are shown in the insets. The scaling pressure is $p_n = k_B T/a^3$. }
\end{figure}

\begin{figure}
$\begin{array}{c}
\includegraphics[scale=0.6, clip]{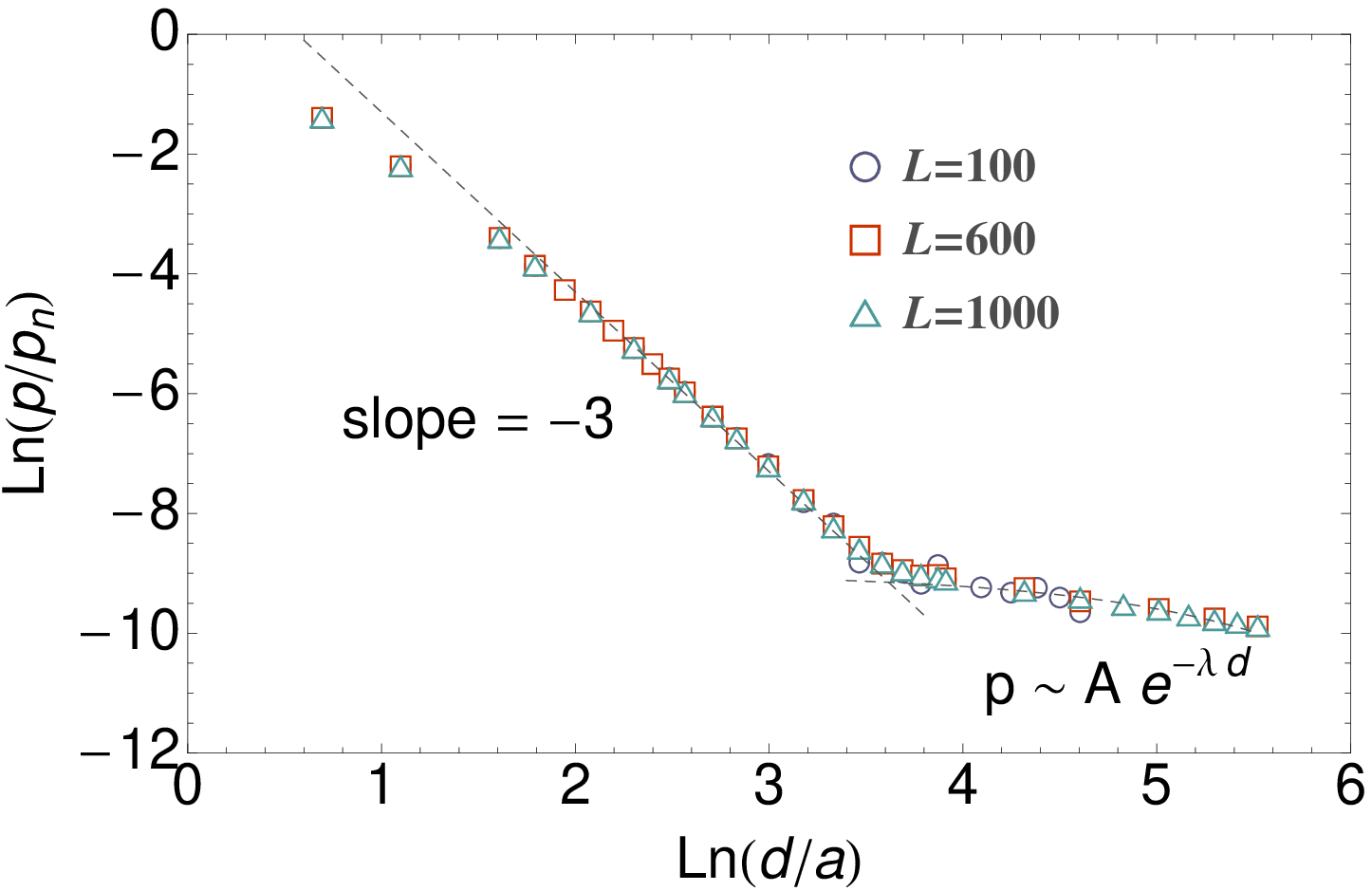}
\end{array}
$
\caption{\label{fig:5}  Simulation results (symbols) of  the $p-d$ dependence at the longer range of $d$'s.  The sizes of the membrane are shown in the labels. 
The other simulation parameters are $k_BT = 10.0$,  $\kappa = 1.0$,  and $a = 1.0 $.  The scaling pressure is $p_n = k_B T/a^3$. }
\end{figure}

In most simulation runs,  at least $8\times 10^4$ MC time steps were performed.  A Monte Carlo time step is defined as  the number of Monte Carlo updates divided by the number of grid points of the membrane.  The first 3000 time steps were omitted for thermalization.  The calculated statistical errors of a few percent of the energy estimators  are obtained. A typical simulation result for heating of  a membrane is shown in Figure \ref{fig:3}.  The size of the membrane in this figure is 120 by 120, where $a = 1.0$,  $d = 5.0$, and $\kappa =1.0$.  To give a sense about the errors on the average energy, the measured internal energy density of Figure \ref{fig:3} at $k_B T/\kappa \approx 20$ is $\bar{\Ee} = 9.080\pm 0.009$.

The pressure-distance relation for the separation ranging from $0.1-7.0$ is shown in Figure \ref{fig:4}. 
The simulation parameters are  $k_BT = 10.0$,  $\delta d = 0.1$,  and $a = 1.0 $, while the bending moduli are $\kappa = 0.1 k_BT$, $5k_BT$,  and $20k_B T$, respectively. 
The size of the membrane shown are $L\times L = 50\times 50$, $600\times 600$, and $1000\times 1000$. 
From these log-log plots, it is clear that the 
relation $p\sim 1/d $ is applicable within a small range $d$,  while there is a transition to $p\sim 1/d^3$ as the 
separation distance increases. The transition from $1/d^3$
to $1/d$-dependence can be estimated  from Eq.(\ref{eqn:13}) as
\begin{equation}
 d < d_{\text{tran}} = \sqrt{\frac{2 k_BT a^2}{\kappa}},
 \label{eqn:29}
\end{equation}
which for the choice of parameters $k_B T = 10$, $\kappa = 1.0$, $a=1.0$  the corresponding transition length is 
$d_{\text{tran}} \approx 4.5$ or about five times
the spacing between molecule ($ a= 1.0$).  For more realistic values of $\kappa \approx 20 k_B T$ and $a \approx 8$ \AA$\,$ \cite{Goetz98}, the transition length
is $d_\text{tran} \approx 2.5\,$\AA. 

Is the transition from $1/d$ to $1/d^3$-pressure law a result of finite size effects?
Our MC results in Figure \ref{fig:4}  for the membrane as large as $1000\times1000$ indicate otherwise. 
As evident, the smaller membrane of size $L\times L = 50\times 50$ exhibits the same transition length $d_\text{tran}$  to those 
of much larger ones.  Since in real biological membranes
the number of possible excitation modes are perhaps much larger than in our model,  the transition from the ideal gas $1/d$ to $1/d^3$ pressure law could possibly
represent some new physics---the exploration of this is deferred to future work.  Our results are easily interpreted within the context of the free energy of a tightly confined membrane. At small $d/a$, the bending
energy contribution to the free energy is negligible compared to the entropic contribution. The suppression of the elastic effects hence leads to the limiting pressure law
of the ideal gas. 

Nevertheless, it should be noted that the transition length $d_\text{tran}$  depends
on the bending modulus, temperature, and intermolecular spacing as indicated in Eq.(\ref{eqn:29}). 
The pressure law transition from $1/d$ to $1/d^3$ dependence  may have some important implication in the interactions among living cells, potentially opening 
a new avenue in reevaluating the conventional understandings about how biological cells mechanically interact.

A larger range of the $p-d$ dependence is shown in Figure \ref{fig:5}.  The sizes of the membrane, which are $L$ by $L$, are shown in the subfigures. At large $d$  the pressure follows an exponential decay relation of type $p \sim A \exp(- \lambda d)$ resulting in the log-log plots of the form $y= D-\lambda \exp(x)$ 
where $x\equiv \ln d$,  $y \equiv \ln p$, and $D \equiv \ln A$.  This exponential decaying regime  has not yet been confirmed 
by any simulations or theory thus far, despite a speculation in Ref. \cite{Janke86}. An interesting conclusion of this result is that the Helfrich entropic force is not \emph {really} as long-ranged as previously believed.

\section{Conclusions}
In summary, we conclude that Freund's \cite{Freund13} conclusions are correct for short inter-membrane distances and in that regime, his result is a major modification of the well-accepted entropic force law due to Helfrich \cite{Helfrich78}. However, Helfrich is correct for intermediate distances and finally, for large separations, the entropic force decays exponentially.  At the time of writing this manuscript, we became aware of a pre-print by T. Auth and G. Gompper, who (at least for short and intermediate membrane separations) have reached similar conclusions as us.

The physical consequences of the modification of the entropic force law between membranes remains an open problem and is expected to be an interesting avenue for future research.

\begin{acknowledgments}
P. Sharma gratefully acknowledge helpful discussions with Professor Ben Freund and his encouragement to pursue this work. Y. Hanlumyuang thanks  Dr. Xu Liu and Professor Aiichiro Nakano for answering several questions regarding parallel computations,  and Dr. Dengke Chen for countless insightful discussions.
\end{acknowledgments}


\begin{thebibliography}{10}

\bibitem{Freund13}
L.~B. Freund.
\newblock {\em PNAS}, 110(6):2047, 2013.

\bibitem{Helfrich78}
W.~Helfrich.
\newblock {\em Zeitschrift f\"{u}r Naturforschung}, 33(3):305, 1978.

\bibitem{Helfrich84}
W.~Helfrich and R.~M. Servuss.
\newblock {\em Nuovo Cimento C}, 3(1):137, 1984.

\bibitem{Ninham70}
R.~W. Ninham and V.~A. Parsegian.
\newblock {\em J. Chemical Physics}, 53:3398, 1970.

\bibitem{Rand89}
R~.P. Rand and V.~A. Parsegian.
\newblock {\em Biochim. Biophys. Acta}, 988:351, 1989.

\bibitem{Lipowskychapter}
R.~Lipowsky and E.~Sackmann.
\newblock {\em Handbook of Biological Physics, Vol. 1}.
\newblock {Elsevier, Amsterdam, 1995.}

\bibitem{Israel92}
J.~N. Israelachivili and H.~Wennerstrom.
\newblock {\em J. Phys. Chem.}, 96:520, 1992.

\bibitem{Milner92}
S.~T. Milner and D.~Roux.
\newblock {\em J. Phys. I France}, 2:1741, 1992.

\bibitem{Lipowsky86}
R.~Lipowsky and S.~Leibler.
\newblock 56:2541, 1986.

\bibitem{Lipowsky89}
R~Lipowsky and B~Zielinska.
\newblock {\em Phys. Rev. Lett.}, 62:13, 1989.

\bibitem{Janke86}
W.~Janke and H.~Kleinert.
\newblock {\em Phys. Lett. A}, 117(7):353, 1986.

\bibitem{Gouliaev98}
N.~Gouliaev and J.~F. Nagle.
\newblock {\em Phys. Rev. E}, 58:881, 1998.

\bibitem{Kleinert99}
H.~Kleinert.
\newblock {\em Phys. Lett. A}, 257:269, 1999.

\bibitem{Sharma13}
P.~Sharma.
\newblock {\em PNAS}, 110(6):1976, 2013.

\bibitem{Kittel80}
C.~Kittel and H.~Kroemer.
\newblock {\em Thermal Physics}.
\newblock {W. H. Freeman and Company}, 1980.

\bibitem{Kroll89}
G.~Gompper and D.~M. Kroll.
\newblock {\em Europhys. Lett.}, 9:58, 1989.

\bibitem{David90}
F.~David.
\newblock {\em J. de Phys.}, 51:C7--115, 1990.

\bibitem{Netz95}
R.~R. Netz and R.~Lipowski.
\newblock {\em Europhys. Lett.}, 29:345, 1995.

\bibitem{Nakano}
A.~Nakano.
\newblock {http://cacs.usc.edu/education/cs653.html}, 2010.

\bibitem{Heermann91}
D.~W. Heermann and A.~N. Burkitt.
\newblock {\em Parallel Algorithms in Computational Science}.
\newblock{Springer-Verlag}, 1991.

\bibitem{Feynman39}
R.~P. Feynman.
\newblock {\em Phys. Rev.}, 56:340, 1939.

\bibitem{Goetz98}
R.~Goetz and R.~Lipowsky.
\newblock {\em J. Chem. Phys.}, 108(17):7397, 1998.

\end{thebibliography}
\end{document}